# An optical mechanism for aberration of starlight

**Robert A. Woodruff**

Center for Astrophysics and Space Astronomy, Astrophysics Research Laboratory, University of Colorado, 593 UCB, Boulder, Colorado 80309-0593, USA.

raw@colorado.edu

## ABSTRACT

We present a physical-optics based theory of the physical mechanism for aberration of starlight. We apply non-relativistic and relativistic theories for wavefront image formation and include the effects of optically transmitting media within the sensor. We show that the sensors imaging properties combined with finite velocity of light fully accounts for aberration. That is, the influence of the moving sensor on the incident wavefront from the star fully explains aberration. Our treatment differs from all previous derivations because we include wavefront-imaging physics within the sensor model. Our predictions match Earth-sensor based measurements, but differ at larger sensor speeds from predictions of the special relativistic-based theory. While experimental uncertainty resulting from the low Earth-orbital velocity prevents experimental confirmation of the special relativistic model of aberration, we find that Earth-based sensors containing refractive optical media could experimentally differentiate between these competing descriptions and, in addition, yield an independent test of time dilation. We derive and present the details of such an experiment.

## 1. INTRODUCTION

In 1728, Bradley [1, 2] detected aberration of starlight, the seasonal variation of the measured angular position of stars on the celestial sphere, using an Earth-borne sensor with a simple telescope. The ensuing three centuries have seen many explanations for stellar aberration proposed as the understanding of the physical properties of light evolved. Liebscher [3] reviews many of these explanations. None includes the optical properties of the measuring optical sensor. All, except Fresnel [4-6], incorrectly treat light as a particle (i.e., a "rain drop" model [7]) not as an electro-magnetic wave. While Fresnel treats light propagation as wave phenomenon, well before Maxwell [8] established such a model, his treatment requires aether: a construct dismissed after the null results of the Michelson-Morley experiments [9, 10]. Special relativity, which infers that the incident wavefronts tilt relative to the moving sensor due to time dilation in its frame [11], provides the most successful description of aberration [7].

The measurable of aberration of starlight, the aberration angle, is the angular deviation of the celestial position of the star relative to the undeviated viewing direction. The Unaberrated Viewing Direction, defined as the measured stellar coordinates of the star relative to the sensor when the sensor motion is solely parallel to the line between the sensor and the center of the star, is parallel to a straight line between the sensor and the center of the star. Based upon sensor-velocity limited Earth-based sensor measurements, aberration manifests itself with five defining characteristics.

A. The aberration angle varies with a one-year temporal period.

B. The aberration angle is a maximum when the Earth orbital velocity is orthogonal to the undeviated viewing direction and is zero if the Earth orbital direction is parallel (or anti-parallel) to the Unaberrated Viewing Direction.

C. When measured with this Earth-based sensor, the maximum aberration angle is $i \approx v/c$, where $v$ is the mean tangential orbital velocity of the Earth relative to the Sun and $c$ is the speed of light in air (or effectively in vacuum). 1

D. The magnitude and phase of the aberration angle depends on the direction and speed of the motion of the sensor relative to the stellar viewing direction. [7] [13]

E. The effect is completely insensitive to motion of the source star relative to the sensor. [13]

---

1 Throughout this paper we apply the internationally accepted values, in units of kilometer/second, of $c$ = 299,792.458 and $v$ = 29.7846918, using 0.01720209895 radians per day as the value for the Gaussian constant. [12] Thus, we use $\beta = v/c$ = 9.9351*10-5 radians = 20.4926226 arc seconds. In air ($n$ = 1.000288), the speed of light is 299,706.059 km/sec, so an air filled sensor would measure $\beta$ = 20.49853 arc seconds.

Clearly, a complete theory of starlight aberration must include the known properties of light propagation and the physical influence of the sensor on the measurement neither of which is properly included in the special relativity based aberration model. In this paper, we derive a modern theoretical model of the aberration of starlight that includes known wave properties of light propagation and of imaging. We impose two fully accepted physical assumptions related to propagation of light: (1) light propagates as a transverse electromagnetic wave [15] and (2) light propagates at a finite speed [7]. Our model fully explains all five defining characteristics of the measurements without postulating aether, motion-induced wavefront tilt, time dilation, or any non-classical physics concept. Combining the action of the sensor on the incident plane wavefront with time delay due to finite speed of light, we give a complete description of stellar aberration and develop insight into the physical mechanisms of light propagation and detection with a sensor moving even at relativistic velocities.

Section 2.A. presents the geometry used in our analysis. Section 2.B. presents both classical and special relativistic theories of image formation as they apply to aberration measurements. Sections 2.C.1. and 2.C.2. apply these theories to derive aberration functional dependencies for a vacuum (and effectively air) filled sensor as viewed by two different observers: an observer fixed relative to the sensor and an observer who views the sensor in uniform motion. These derivations include all applicable special relativistic effects. Section 2.C.3. relates these results to Bradley's Earth-based experiment. Section 2.D. expands the theory to include sensors filled with glass and develops tools used in Section 3. to define an experiment that would test the predictions. For completeness, Section 2.E. presents the currently accepted theory of aberration based on special relativity.

## 2. ABERRATION OF STARLIGHT BASED ON WAVE PROPAGATION MODEL

### A. Adopted geometry

**Figure 1** presents the overall geometry of the measurement and defines four axes/directions. A distant star emits light from its surface as spherical wavefronts. These expand in all directions as they propagate through space and over a great distance, where they reach the observer, effectively become planar of infinite extent. The incident light propagates normal to its wavefront [15] in the direction of the Poynting vector $\vec{\mathrm{S}} = \frac{c}{4\pi}(\vec{\mathrm{E}} \times \vec{\mathrm{H}})$. The instantaneous electric field vector $\vec{\mathrm{E}}$ and magnetic field vector $\vec{\mathrm{H}}$ lie in a plane tangent to the wavefront. Thus $\vec{\mathrm{S}}$, the cross product of these two vectors is always perpendicular to the wavefront. An observer with sensor moves uniformly with velocity $\vec{\mathrm{V}}$ at a fixed angle relative to the plane of these plane wavefronts. The sensor consists of an ideal well-corrected imaging optical telescope connected rigidly to an accurate position-sensing detector. The telescope forms an image of the distant star onto the detector as a point spread function. The detector lies in an infinite plane perpendicular to the optical axis intersecting the optical axis at the paraxial focal point of the telescope. The rigidly connected detector and telescope form the sensor. They move as a solid unit. The detector measures the location of the centroid of the imaged light distribution. We designate the centroid of the stellar image on the detector as F $(t_0)$ and the physical center of the telescope aperture as V $(t_0)$. As our notation indicates, these positions are functions of time, t. The sensor optical axis is the line passing through points V $(t_0)$ and F $(t_0)$ at any arbitrary time $t_0$.

We define a relative Cartesian coordinate system (x, y, z) and orient the star's unaberrated position along the positive y-axis. We refer to this direction as the Unaberrated Viewing Direction. The incident Poynting vector is parallel to the y-axis. The incident plane wavefronts are perpendicular to the y-axis, and therefore parallel to the x-z plane. The aberrated position of the star is on a line in object space between the center of the sensor aperture and the measured location of the star on the sky, the Aberrated Viewing Direction, labeled as the y′-axis. If the sensor optical axis is oriented parallel to the y-axis, the acute angle in the x-y plane between the y′-axis and the optical axis is the aberration angle $i(\theta)$. The velocity vector $\vec{v}$ of the sensor, which lies in the x-y plane, makes an angle $\theta$ relative to the +y-axis and an angle of $(\pi/2)-\theta$ relative to the incident plane wavefronts. Thus, the y′-axis lies in the x-y plane as do the components of the sensor velocity parallel to the incident wavefronts that is $v_x = v\sin(\theta)$ and the component perpendicular to the incident plane wavefronts that is $v_y = v\cos(\theta)$.

### B. Wavefront based image formation

#### *1. Classical non-relativistic theory of image formation*

An optical imaging system, e.g., an imaging telescope, with a finite aperture diffracts and images incident wavefronts. Following Goodman [16], who presents the well-known non-relativistic theory for image formation, the complex field across a plane immediately behind the aperture $U_l'(x,z)$ relates to the complex field incident on a plane immediately in front of the aperture $U_l(x,z)$ by $U_l'(x,z) = t_l(x,z)U_l(x,z)$ where $t_l(x,z)$ is the complex phase transformation function of the imaging aperture. The optic within the aperture provides the imaging phase transformation. In our case

$$U_l(x,z) = C\,, \tag{1}$$

describes the incident non-tilted plane wave parallel to the $(x,z)$ plane, where C is a real constant. The optic converts $U_l(x,z)$ into converging spherical wavefronts $U_l'(x,z)$ that image at the paraxial focal point of the optic. This is the far-field Fraunhofer condition for image formation. Thus,

$$U_l'(x,z) = C' \exp\left[-\frac{ik(x^2+z^2)}{2f}\right], \tag{2}$$

where $k = \frac{2\pi}{\lambda}$ is the wave number, $\lambda$ is the wavelength of light, *f* is the paraxial focal length of the optic, and $C'$ is a real constant. Thus, the phase transformation function of the optic becomes

$$t_l(x,z) = \frac{C'}{C} \exp\left[-\frac{ik(x^2+z^2)}{2f}\right]. \tag{3}$$

Equation (2) shows that the optic converts the incident plane wavefronts into spherical wavefronts that converge to form an image, a Point Spread Function (PSF), at the focal point of the telescope. By Equation (3) the telescope optics introduce a quadratic phase shift to the portion of incident wavefronts contained within the physical extent of the telescope aperture.

The PSF is the square of the Fourier Transform of the complex Pupil Function $P_l(x,z) = t_l(x,z)$ inside of the shape of the aperture, Σ in **Figure** 1. If the aperture were circular, unobscured, with ideal quadratic imaging phase shift, the resultant Point Spread Function is the well-known Airy pattern [16]. This PSF has circular symmetry about its centroid. The portions of the planar wavefronts that lie outside the finite extent of Σ, also diffract, by Babinet's principle, as spherical wavefronts, but experience no phase shift from the telescope. They focus at infinity and do not contribute to the image of the star.

### *2. Special relativistic theory of image formation*

Hillion [17] developed a relativistic theory of scalar and vector diffraction by planar apertures. He applied the theory to find far-field imaging solutions, i.e., Fraunhofer-like solutions, for diffraction by two distinct aperture geometries: rectangular and circular. He found in both cases that the non-relativistic Fraunhofer solution applies in the relativistic case if both the point of observation and the source do not have large transverse coordinates with the proviso that relativistic length contraction modifies the aperture shape.

Our treatment complies with Hillion's restrictions that the point of observation and the source do not have large transverse coordinates. Therefore, we apply non-relativistic Fraunhofer theory in the relativistic case with suitable accounting of aperture geometry changes due to relativistic effects. We note that the location of the centroid of the image is unaffected by relativistic effects, since the aperture shape remains symmetrical.

Modifying Equation (3) to a relativistic condition the phase transformation of an optic which is moving with uniform velocity $v = \beta c$ in the *x*-direction, including wavelength change due to the Transversal Doppler Effect from $\lambda$ to $\lambda_{shift}$ where $\lambda_{shift} = \gamma\lambda$, [18] becomes

$$t_l(x,z) = \frac{C'}{C} \exp\left[-\frac{ik(x^2+z^2-\beta^2x^2)}{2\gamma f}\right]. \tag{4}$$

A circular aperture optic of radius $r_0$ in its rest frame becomes elliptical with semi major axis $r_0$ and semi minor axis $\frac{r_0}{\gamma}$ along the direction of motion, the x-axis, due to relativistic length contraction, where $\gamma = \frac{1}{\sqrt{(1-\beta^2)}}$ is the Lorentz factor. The PSF intensity acquires elliptical symmetry centered about its centroid. With the optic well corrected to be free of internal wavefront aberrations, the centroid position remains centered on the chief ray.

## C. Aberration of starlight with sensor using vacuum-filled telescope

We now consider two observational geometries: (1) the sensor motion directly toward the star with $\theta = 0$ and (2) the sensor motion perpendicular to the line from the sensor to the star with $\theta = \pi/2$. For both observational geometries, we consider aberration of starlight measured by two different observers: (a) an observer in the frame of reference of the sensor and (b) an observer in a frame fixed in the defined Cartesian coordinate system who views the sensor moving with uniform velocity. We refer to the first (a) as the "sensors frame" and the latter (b) as the "coordinate system frame".

*1. $\theta$ = 0 degree viewing geometry*

The $\theta = 0$ observation is illustrated in **Figure 2**. The sensor views plane wavefronts of infinite extent from the distant star. The sensor has zero velocity in a direction parallel to, i.e., along the incident wavefronts. Any non-zero component of its velocity *v* is strictly perpendicular to the incident wavefront. Thus, $v_x = 0$ but $v_y$ may be non-zero in the coordinate system frame. The incident Poynting vector, parallel to the y-axis in **Figure 1**, is parallel to any non-zero sensor velocity component.

*(a) In the sensor frame of reference for a $\theta$ = 0 degree viewing geometry*

In the frame of reference of the sensor, the source moves toward from the sensor along the sensor optical axis. The incident plane wavefronts remain planar and perpendicular to the optical axis. The phase shifted converging wavefronts move exactly along the optical axis of the sensor and image on-axis at the focal point centered on the detector. The Aberrated Viewing Direction and the optical axis of the sensor are parallel. The measured centroid is unchanged by the motion. The wavelength is Doppler shifted from $\lambda_0$ to $\lambda_0\sqrt{\frac{(1-\beta)}{(1+\beta)}}$. No relativistic effects occur within the sensor because this sensor-fixed observer has no motion relative to the sensor. The sensor records zero aberration satisfying Characteristic B of Section 1. This is the measured result when the Earth orbital velocity is parallel to the line from the Earth to the star. Since source relative motion will not change the geometry of the incident plane wavefronts, the result is independent of source motion: Characteristic E of Section 1.

*(b) In the coordinate system frame for a $\theta$ = 0 degree viewing geometry*

An observer in the frame fixed relative to the coordinate system sees the sensor moving with uniform velocity perpendicular to the incident wavefronts. Thus, the imaging converging wavefronts move exactly along the optical axis of the sensor and image at the center of the detector. The Aberrated Viewing Direction of the measurement and the optical axis of the sensor are parallel. Relativistic length contraction affects the converging wavefront and the distance from the vertex to the image plane. The departure from a plane normal to its vertex given in Equation (2) contracts from $\frac{\sqrt{(x^2+z^2)}}{4f_0}$ to $\frac{\sqrt{(x^2+z^2)}}{4f_0\gamma}$. The focal length in Equation (2) also contracts $f_0$ to $\gamma f_0$. Since these cancel in Equation (2), the phase shift of Equation (2) applies. The wavelength is Doppler shifted from $\lambda_0$ to $\lambda_0\sqrt{\frac{(1-\beta)}{(1+\beta)}}$. The axial distance from the optic to the detector contracts from $f_0$ to $\gamma f_0$. This matches the change in optic focal length. The image is in focus. We conclude that the angular position of the star is unchanged and that starlight aberration is zero, satisfying Characteristic B of Section 1 for $\theta = 0$ degrees. Since source relative motion will not change the geometry of the incident plane wavefronts, the result is independent of source motion: Characteristic E of Section 1.

*2. $\theta$ = 90 degree viewing geometry*

The $\theta = 90$ degree observation is illustrated in **Figure 3**. As before, the sensor views plane wavefronts from a distant star. The sensor motion is parallel to the plane of the incident wavefronts. The sensor has zero velocity in a direction perpendicular to the incident wavefronts. Thus, $v_y = 0$ and $v_x \neq 0$.

*(a) In the sensor frame of reference for a $\theta$ = 90 degree viewing geometry*

As before, the finite telescope aperture diffracts the incident wavefronts introducing a quadratic phase shift to the portion of the wavefronts that lie within the telescope aperture converting these into spherical wavefronts. These converge to image at the position that the telescope focal point *was at the instance when* the wavefront encountered the limiting diffracting aperture Σ of the telescope. (As before, the planar wavefronts that lie outside this aperture also diffract and focus at infinity.) Each wavefront, respectively, encountering the aperture at position $V(t_1)$ converges to position $F(t_1)$ for each arbitrary time $t_1$. The diffracted Poynting vector is coincident to the wavefronts velocity vector, orientated at 90 degrees to the sensor velocity vector. Thus, to an observer in the sensor frame the center of curvature of the converging wavefronts move parallel to the optical axis of the sensor. The wavefront images at the exact position the detector *was at the instance when* the aperture diffraction occurred. We arbitrarily designate this time as $t_0$.

Now applying our second assumption that light has a finite velocity *c*, each imaged wavefront will reach the image plane a finite time $t_f$ after $t_0$. Thus, $t_f = f/c$ where *f* is the paraxial focal length of the sensor and *c* is the average speed of light in the medium between the telescope imaging optic and the detector. While the light converges to the detector plane, the sensor with its detector translates parallel to the detector plane by a distance $x = vt_f$. The optic forms the convergent imaging wavefront of the Point Spread Function *when the detector is at position* F ($t_0$), but the imaging wavefront images at the detector *when the detector is at a different position*, F ($t_0+t_f$). To the sensor, the star appears to be off-axis by an angle, *i*, satisfying

$$\tan(i) = v/c\,. \qquad (5)$$

The sensor translates in the x-direction during the temporal interval $t_f$ from when a given wavefront encounters the aperture to the instance that it reaches the detector forming an image of the star. In the sensor reference frame, the image forms at coordinates ($-f\tan(i)$, *-f*, 0). Thus, the star appears off-axis to the sensor by the angle given in Equation (5). The observer senses aberration of starlight. The diffracted converging wavefronts appear to translate perpendicular to the optical axis shearing as they converge to form the stellar image. The wavefronts do not tilt. The observer fixed relative to the sensor senses no relativistic effects. The sensor detects the non-relativistic PSF. If $v << c$, $i \approx v/c$ as expected and if $v = c,\ i = 45$ degrees.

Planar wavefronts arrive continuously at the imaging aperture. For any arbitrary time interval *δ*, a wavefront arriving at the imaging aperture at time $t_0 + \delta$ will reach the detector at time $t_0 + \delta + t_f$. Thus, a continuous image displaced by the sensor motion that occurred during time $t_f$ results. The Aberrated Viewing Direction tilts with respect to the optical axis of the sensor. We conclude that the position of the star has changed and therefore we detect non-zero aberration of starlight. This geometry occurs when the Earth orbit direction is perpendicular to the line from the Earth to the star.

*(b) In the coordinate system frame for a θ = 90 degree viewing geometry*

For the *θ = 90* degree viewing geometry, an observer fixed relative to the coordinate system of **Figure 1** views the sensor moving uniformly in the x-direction. As was the case for the observer fixed relative to the sensor, the sensor views plane wavefronts from a distant star and the sensor motion is parallel to the plane of the incident wavefronts. However, in this case, relativistic effects modify the aperture and optics phase shift. A circular aperture would become elliptical and a spherical wavefront would become elliptical. Because these relativistic effects are symmetric varying functionally with $\beta^2$, image symmetry and image centroid location are the same as for the non-relativistic sensor frame measurement. The Fraunhofer approximation PSF, modified by relativistic changes in the aperture shape, applies in this relativistic case based on Hillion [17].

The axial separation between the optic and detector is unchanged by relativistic effects because all motion relative to the coordinate system observer is normal to the axial direction. Since light has a finite velocity *c*, the imaged wavefront does not reach the image plane until a finite time $t_f = f/c$ after encountering the imaging aperture. While the light is converging to the detector plane, the sensor with its detector translates parallel to the detector plane by a distance $x = \frac{vt_f}{\gamma} = \frac{vf}{\gamma c}$. The imaging convergent wavefront forms the Point Spread Function *when the detector is at position* F ($t_0$), but the imaged wavefront arrives at the detector when *the detector is at a different position*, F ($t_0+t_f$). The star appears to be off-axis in the sensor by an angle, *i*, satisfying $\tan(i) = \frac{v}{\gamma c}$, which differs from the sensor frame measurement given in Equation (5) by the Lorentz factor $\gamma$.

Planar wavefronts arrive continuously at the imaging aperture. For any arbitrary time interval $\delta$, a wavefront arriving at the imaging aperture at time $t_0 + \delta$ will reach the detector at time $t_0 + \delta + t_f$. The sensor motion that occurs during time $t_f$ displaces the resulting continuous image. The Aberrated Viewing Direction tilts with respect to the optical axis of the sensor. We conclude that the position of the star has changed and detect non-zero aberration of starlight.

The converging wavefronts do not shear laterally and do not tilt, but behave exactly as they did for the $\theta = 0$ case. In the frame of the moving sensor, the diffracted converging wavefronts appear to translate as they converge to form the stellar image and no relativistic effects occur. Due to relativistic effects, the coordinate system observer detects a change in image profile, but no change in image centroid location because the PSF maintains symmetry. Neither of these observers senses a tilt in the incident plane wavefront. Thus, the Aberrated Viewing Direction tilts relative to the optical axis of the sensor. Because of relativistic effects, the two observers measure the different values for the aberration angle.

*3. Bradley's experiment: Sensor on Earth surface*

We generalize by designating the component of the sensor velocity in a direction parallel to the incident plane wavefronts as $v_x$. Modifying Equation (5), the stellar aberration tilt angle becomes in the sensor frame of reference

$$i = \arctan(v_x/c) \text{ or equivalently } \tan(i) = (v_x/c). \quad (6)$$

To compare to Bradley's experimental results, we now consider measurements in the sensor frame with a sensor filled with air on the surface of the Earth. We assume the Earth motion around the Sun to be circular with radius $R_E$ lying in the X-Y plane with tangential orbital speed $v$ relative to the Sun and place the origin of the X-Y system at the center of the Sun. The unaberrated position of the observed star lies along the +Y-axis. The Earth orbital location is $(X_E, Y_E)$ and Earth orbital velocity components $(v_x, v_y)$. Thus,

$$X_E(\theta) = R_E \cos(\theta), Y_E(\theta) = R_E \sin(\theta), v_x(\theta) = v\sin(\theta) \text{ and } v_y(\theta) = v\cos(\theta). \quad (7)$$

Equation (6) becomes

$$\tan(i(\theta)) = (n_{TEL} v_x/c) = n_{TEL} v \sin(\theta)/c = n_{TEL} \beta \sin(\theta), \text{ where } \beta = (v/c). \quad (8)$$

Aberration expressed in Equation (8) has a one-year period consistent with Characteristic A of Section 1. The measurement depends only on the sensor velocity component parallel to the incident planar wavefronts: Characteristic D of Section 1. The orientation of these incident wavefronts is independent of source motion thus the measurement is totally insensitive to motion of the source: Characteristic E of Section 1.

By Equation (8), the maximum aberration angle occurs when $\theta = \pi/2$ and satisfies $\tan(i) = n_{TEL}\beta$. The angle is zero when $\theta = 0$ in agreement with Characteristic B of Section 1. This Earth-based observation with $\beta << 1$ would measure $i \approx \beta$, (taking $n_{TEL} = 1.0$) which agrees with Characteristic C of Section 1. Therefore, the wave imaging theory predicts all five aberration characteristics outlined in Section 1.

From special relativity, the sensor velocity cannot exceed the velocity of light, so $v \le c$. Thus, the aberration angle of a vacuum-filled sensor is restricted to $i \le 45$ degrees. If the sensor velocity is the speed of light $\beta = 1$, Equation (8) becomes $\tan(i(\theta)) = 1$ and the aberration angle becomes exactly 45 degrees. The aberration angle in an air-filled sensor is restricted to $i \le 45.0082494$ degrees, using $n_{air}$ = 1.000288. This value is 29.7 arc seconds larger than for a vacuum-filled sensor value.

## D. Aberration of starlight with sensor using telescope filled with glass

Aberration depends on the speed of light and hence on the refractive index within the sensor, so we consider the measurement in the sensor frame of reference with a sensor filled with non-vacuum optical media. **Figure 4** illustrates starlight aberration as measured with a telescope constructed of a block of glass (or other rigid transparent optical medium) of refractive index, *n*, with the detector at the rear surface of the glass block. The first surface is a convex refracting surface with a sharp edge forming the telescope aperture stop. This surface will form a monochromatic diffraction-limited PSF image of a distant star at the plano rear surface. The focal length in the optical medium is *f*, so the focal length in vacuum is *f/n*. The distance from the vertex of the front surface, V ($t_0$), to the focal point, F ($t_0$), measured on a line normal to the rear surface is *f*. The line between V ($t_0$) and F ($t_0$) is the optical axis of the sensor. The rear surface of the glass block is normal to the optical axis.

As before, the maximum aberration will occur when the incident plane wavefronts from the star are parallel to the velocity vector of the sensor. The propagation direction of the wavefronts is then parallel to the optical axis of the sensor. The optically powered front surface will convert each incident plane wavefront into a spherically converging wavefront. The converging wavefront will form an image of the star at the position where point F *was* when the wavefront refracted by the front surface. We designate this image position as F ($t_0$). The wavefront requires a time interval, $\delta t = nf / c$ to reach the detector plane, where $c / n$ is the average phase velocity of light in the medium. The sensor moves parallel to the detector plane a distance $x = v_x \delta t = nfv_x / c$ during the time interval for the light to reach the detector plane as an image.

Therefore, the sensor measurement concludes the source is off-axis. Its angular position on the sky is "aberrated" by an angle, *r*, measured within the optical medium where $\tan(r) = x/f = nv_x / c$. By Snell's law, the corresponding angle in object space, *the aberration angle i*, satisfies $n_{object} \sin(i) = n \sin(r)$, where $n_{object}$ is the refractive index of object space. Thus,

$$n_{object} \sin(i) = n \sin(\arctan(nv_x / c)). \qquad (9)$$

As before, $\theta$ is the angle between the instantaneous direction of motion of the Earth in its orbit to the Earth/star line of sight. The component of orbital velocity *v* projected onto the plane of the incident wavefront is $v_x = v \sin(\theta)$. Thus, the aberration angle measured by an Earth-based sensor varies over the year and satisfies

$$n_{object} \sin\{i(\theta)\} = n \sin\{\arctan(n\beta \sin(\theta)\}, \text{ where } \beta = v/c. \qquad (10)$$

For the Earth-based stellar observation with $\beta << 1$ and $n_{object} = 1$, the quantity $n\beta \sin(\theta) << 1$, so

$$i(\theta) \approx \frac{n^2 \beta \sin(\theta)}{n_{object}}. \qquad (11)$$

Thus, the maximum aberration geometry occurs at $\theta = \pi/2$ and is

$$i(\pi/2) \approx n^2 \beta \text{ if } \beta << 1 \text{ where we take } n_{object} = 1. \qquad (12)$$

The aberration angle varies from zero at $\theta = 0$ to a maximum value of $i(\theta = \pi / 2) \approx n^2 \beta$.

For larger sensor speeds, Equation (10) requires $\sin\{i(\theta)\} \leq 1$, so if $n_{object} = 1.0$, we find that $n \sin\{\arctan(n\beta \sin(\theta)\} \leq 1$. If the measurements are made at the maximum aberration geometry, $\theta = \pi/2$, at the wavelength of 656.28 nm (H$\alpha$), the aberration angle is real for $\beta < 0.855$ for a water-filled telescope ($n$ = 1.331152) and $\beta < 0.649$ for a fused silica telescope ($n$ = 1.456369). Larger values of $\beta$ will result in a complex value for the aberration angle with a 90-degree real part. If the telescope and object space are vacuum-filled, Equation (10) yields $\tan(i(\theta)) = \beta \sin(\theta)$. Special relativity requires that $\beta \leq 1$. Thus, $i(\pi/2) \leq 45$ degrees that agrees with Equation (5).

By Equation (12), an Earth-based sensor operating at $\lambda$ = 656.28 nm completely filled with various materials would measure the following values (using $n_{object} = n_{AIR}$):

| | Vacuum | Air | Water | Fused silica | Turpentine |
|---|---|---|---|---|---|
| Refractive index ($\lambda$ = 656.28 nm) | 1.0 | 1.000293 | 1.331152 | 1.456369 | 1.472 |
| Maximum aberration angle (arc sec) | 20.48661 | 20.49862 | 36.30157 | 43.45232 | 44.39006 |

## E. Corresponding predictions based on Theory of Special Relativity

The Theory of Special Relativity attributes starlight aberration to time dilation across the sensor aperture causing the incident wavefront to tilt [7]. The theory does not overtly include the imaging properties of the "telescope", but instead uses a "rain drop" model similar to the Newtonian corpuscular model of light propagation. It finds that $i(\theta)$ is the solution to

$$\cos(\theta - i(\theta)) = [\cos(\theta) + \beta]/[1 + \beta\cos(\theta)] \text{ , where } \beta = {}^{v}\!/_{c} \, . \tag{13}$$

where $c$ is the speed of light in the medium within the "telescope. (See Jackson [14] Figure 11.1)

Based on Equation (13) the aberration angle is zero when $\theta = 0$, as expected. The maximum aberration angle occurs when $\theta = \pi/2$. Based on Equation (13), the maximum aberration is the solution to

$$\sin(\mathrm{i}({}^{\pi}\!/_{2})) = \beta \, . \tag{14}$$

For an Earth-based stellar observation where $\beta << 1$, the maximum aberration angle is

$$\mathrm{i}({}^{\pi}\!/_{2}) \approx \beta \, . \tag{15}$$

Equation (13) predicts Bradley's experimental result over the yearly Earth orbital period, since $\cos(i(\theta)) \approx 1$ because $\beta = (v/c) << 1$,

$$i(\theta) \approx \beta\sin(\theta) \, . \tag{16}$$

Equation (16) is identical to Equation (8) for an air/vacuum-filled telescope with $\beta << 1$.

By Equation (14) with $\theta = \pi/2$ the aberration angle would become complex as the sensor velocity increases if $n\beta$ were greater than unity. Thus, $\beta \leq {}^{1}\!/_{n}$. If the wavelength is 656.28 nm, the aberration angle is real if $\beta < 0.751$ for a water-filled telescope and $\beta < 0.687$ for a fused silica telescope. At larger values, the angle is complex. A vacuum-filled sensor moving at the speed of light measures an aberration angle of 90 degrees.

Predictions using the two models for air-filled sensors illustrate some differences. No difference occurs for $\theta = 0$ degrees. For $\theta = 90$ degrees, the measurements differ by 12.9 arc seconds when $\beta = 0.05$. Thus, differences are small unless the sensor has large relative velocity. However, at the speed of light, the relativistic-based theory predicts $i(\frac{\pi}{2}) = 90^{o}$ and the wave-based theory predicts $i(\frac{\pi}{2}) = 45^{o}$ a significant difference.

## 3. DEFINE NEW EXPERIMENT TO TEST THEORIES

Both Equation (11) and Equation (16) predict that an Earth-based stellar observation like that of Bradley's air-filled telescope would result in aberration of $\beta\sin(\theta)$. Thus, the effect would be sinusoidal with one-year period with 20.4985 arc seconds half amplitude. Such small observer velocities cannot discern between the two theories. [26] However, adding an optical media with $n > 1$ into the telescope would break this degeneracy.

In the latter half of the 19th century, Klinkerfues [19, 20] and Airy [21] attempted such measurements with conflicting results. Each measured aberration of starlight during an Earth annual orbital period using a telescope partially filled with a liquid, instead of an air-filled telescope as used by Bradley. Airy, using a telescope partially filled with pure distilled water, detected no change from Bradley's air-filled telescope measurement and concluded that our Equation (16) represents aberration applying the Fresnel drag hypothesis. These results contradicted previous experimental measurements and theoretical predictions by Klinkerfues [19], and Hoek [22, 23]. Klinkerfues detected a larger aberration angle using a telescope partially filled with a column of pure turpentine. The literature [7], accepts Airy's result.

We now evaluate Klinkerfues's experiment using the techniques of Section 2. D. and 2. E. A refractive thin lens telescope of focal length $f$ filled only with air will focus the on-axis distant star an axial distance $f$ behind the lens. In this image space, light travels with speed $c$, so the time delay from lens to focus is $t_f = f/c$. If the sensor travels at velocity $v$ normal to the line of sight to the star, assuming $\beta << 1$, the image appears off-axis to the moving sensor by angle $i_f = v/c$ (i.e., Bradley's result).

Using elementary first order paraxial optics, inserting a column of length $d_2$ of medium with refractive index $n_2$ into the space between the lens and image surface leaves the telescope of focal length unchanged. The axial distance between the lens and image increases by $d_2(n_2 - 1)/n_2$ [24]. The space between the lens and image is now shared by the column $d_2$ and a column of air of length $d_1$, where $d_1 = f - d_2/n_2$. Light travels with speed $c$ within medium $d_1$ and speed $c/n_2$ within medium $d_2$. The time delay from lens to focus is $t_d = f/c + \frac{d_2(n_2^2 - 1)}{(cn_2)}$. For the maximum aberration condition when the sensor travels at velocity *v* normal to the line of sight to the star, the image appears off-axis in the frame of the moving sensor by angle, again applying the small angle approximation,

$$i_d = i_f t_d / t_f \ . \qquad (17)$$

**Figure 5**, which is Figure 4 from Klinkerfues's paper [20], details his design approach to define the liquid column. Klinkerfues [20] used $f$ = 457.2 mm, $d_2$ = 203.2 mm, and $n_2$ = 1.472. Thus, Equation (17), predicts he would measure an aberration angle of 27.720 arc sec, which is 35.2% larger than for an air-filled telescope. (We have not corrected for the end windows he used to constrain the liquid.) We note that this agrees to better than 1.3 % with Klinkerfues reported result of apparent motion of 27.365 arc seconds. Klinkerfeus's measurement appears to confirm our model.

However, how do we interpret Airy's null result? We studied Airy's publication [21] and found the description of his experiment so severely lacking in detail that we are unable to reconstruct his results. He states that a more detailed follow-up write-up would be prepared. After an exhaustive literature search, we have been unable to locate its reference. Apparently, references to it do not exist, so we surmise he may not have disseminated the follow-up description.

The major operative difference between these two experiments is the approach in containing the liquid. Airy used a vertical telescope (focal length ~ 670 mm) with an unconstrained upper surface of water (n=1.33) that nearly filled the whole ~ 900 mm length from the two-lens objective to the image surface. Klinkerfues used a transit telescope of focal length 457.2 mm into which was inserted a 203.2 mm column of pure turpentine (n=1.472) whose surfaces were in a complete tube with closed glass end plates (See **Figure 5**). Both of the optically functional surfaces of the turpentine column in the Klinkerfues experiment would apparently remain plano and parallel as well as normal to the optical axis of the telescope. Thus, the glass-ended tube could preserve the geometry of the Klinkerfues experiment, while in Airy's geometry gravity could modify the tilt of the upper surface of water.

Airy does not describe how the tilt angle of the top surface was controlled. Only the lower of the optically functional surfaces of the water column in Airy's experiment would remain normal to the optical axis of the telescope. While both surfaces of water would remain plano (outside of their edge meniscus of the upper surface), the upper surface of water could tilt, due to local gravity, as the telescope tips slightly as it measures the aberration angle. A prism of water could form which would change the measurement and possibly invalidate his conclusions. If this occurred, it would nullify his conclusions.

A repeat of the aberration experiment using a telescope with well-defined optical surfaces is the logical recommendation of this paper. The sensor can be a solid block of glass telescope with modern detectors. Based on Equation (12) a fused silica telescope 500 mm long with the front surface a convex conic and the detector at the plano rear surface would detect an aberration angle of 43.47 arc seconds at $\lambda$=656.28 nm (H$\alpha$) with a readily discernable displacement of 72.4 microns at the image. This design example is a solid cylindrical block of fused silica glass 500 mm long of approximately 30 mm diameter. The front surface is optically figured as a convex ellipsoid of revolution with 156.6804 mm vertex radius of curvature and conic constant of K= -0.471473 (A sphere has conic constant of zero and the conic constant of a parabola is -1.0.) A comparable air or vacuum-filled sensor would detect 20.5 arc seconds with a displacement of 49.7 microns, providing a clear test of the predictions.

## 4. CONCLUSION

### A. New theory of aberration of starlight includes wavefront imaging of sensor

Clearly, any aberration of starlight theory must include physically accurate imaging properties of the sensor. No previous model for aberration has properly included the optical characteristics of the measuring sensor. Therefore, we present a new theory for aberration of starlight based on well-established physical theories of light propagation and imagery, including the finite velocity of light.. This is the primary difference between our approach and all others. All of our assumptions are fully compatible with modern physics including relativistic effects on image formation, time dilation, and length contraction. Our study fully supports the Theory of Special Relativity, but calls into question the current application of special relativity to the aberration of starlight phenomenon, since special relativity does not properly model the imaging aspects of the sensor. Our theory agrees completely with current measurements of aberration, encompassing all five of the measured characteristics. We suggest a straightforward experiment to select between competing theories.

### B. Comparison with special relativity-based theory for aberration

The current special relativity-based model correctly predicts most of the measured variables of slow Earth orbit velocity aberration of starlight. Per Equation (16), the special relativity model predicts Characteristic A (one-year temporal period), Characteristic B (maximum at $\theta$ = 90 degrees and minimum $\theta$ = 0 or 180 degrees, Characteristic C (maximum value of $i \approx v/c$ ), and Characteristic D (magnitude and phase depends on relative motion of sensor).

The following argument shows how the theory could include Characteristic E compliance. The principle of non-absolute motion within the special theory of relativity requires symmetry between uniform relative motions of the source and of the observer. Their individual relative motions should be indistinguishable. The relativity-based aberration model as currently presented [14] [25] does not rigorously include symmetry between source and observer motions. Special relativity notes that tilting of the incident plane wavefront retains Lorentz invariance [14]. Jackson [14] shows that Lorentz invariance of wavefront phase is equivalent to relativistic time dilation, and in addition is equivalent to relativistic addition of sensor velocity using a particle ballistic model to characterize light propagation. However, Jackson's derivation of the relativistic model for aberration omits source velocity when combining velocities. The tangential velocity of the source is not included in the addition of velocities, so the derivation assumes asymmetry between source and sensor motion [14] [25] [13]. Textbooks, like Rindler [25], only treat the case of the sensor moving relative to a fixed star and do not explicitly treat the case of relative velocity between the star and the observer.

However, source velocity is not a contributor to aberration. Eisner [13] shows that aberration does not depend on the velocity of the source relative to the sensor. We conclude that aberration must therefore be solely due to motion of the observer relative to the incident wavefront without implying a preferred frame. That is, relative motion of the source must not modify the wavefront captured by the sensor for otherwise aberration would depend on relative motion of the source.
We note that this assessment is consistent with a possible relativistic explanation that the sensor relative motion tilts the wavefront by time dilation. Thus, the special relativity explanation implies wavefront tilt due to sensor relative motion but not due to source motion. Given this explanation, wavefront tilt based only on sensor relative motion is theoretically sufficient to predict Bradley's result. The time dilation hypothesis effectively removes the dependence on source relative motion. With this understanding, compliance to Characteristic E of the relativistic-based model follows.

Our physical optics-imaging model properly predicts all five characteristics. The wavefronts remain plano until aperture diffraction occurs. Thus tangential source motion simply shears the wavefronts in plane. The incident wavefronts appear unchanged by source motion, so no source-motion dependence occurs.

### C. Proposed experiment to resolve aberration theories

The sole motivation for this paper is to apply modern optical wave theory to the aberration of starlight and see what logically results. We raise a simple, but important, question:  Does stellar aberration result from wavefront tilt? The special relativity model requires the incident wavefronts to tilt, and ascribes the tilt to time dilation. Our model does not require tilted wavefronts, and deduces that aberration of starlight results from laterally sheared untilted imaging wavefronts in the observers frame.

We include effects of non-air or non-vacuum media within the optical path of the sensor. This leads us to propose a modern stellar aberration experiment to test our model as well as the relativistic-based model. Since the relativistic-derived model requires wavefront tilt due to time dilation, we note that our proposed experiment would provide an independent test of time dilation.

Only slow velocity tests of aberration have been possible to date. Other authors, e.g., [26] who compare aberration of starlight predictions using Galilean relativity models to special relativity models, have recognized that Earth velocity based measurements cannot confirm the relativistic model. Sartori concludes, "Since the aberration angle itself is only about 20 seconds of arc, the relativistic correction would be order 10-7 arc seconds, much too small to be detected. Hence aberration measurements provide no additional confirmation of special relativity." On this basis, some have concluded that the special relativistic model of aberration remains unconfirmed experimentally. A repeat of a Klinkerfues/Airy type experiment with a sensor of known optical properties will test the theories and break their degeneracy even at slow Earth orbit velocities. We conclude this experiment would provide insight into relativistic imagery and into special relativity. The experiment would provide an independent test to discriminate between the two views. A solid glass telescope with appropriate would suffice.

**Acknowledgements**

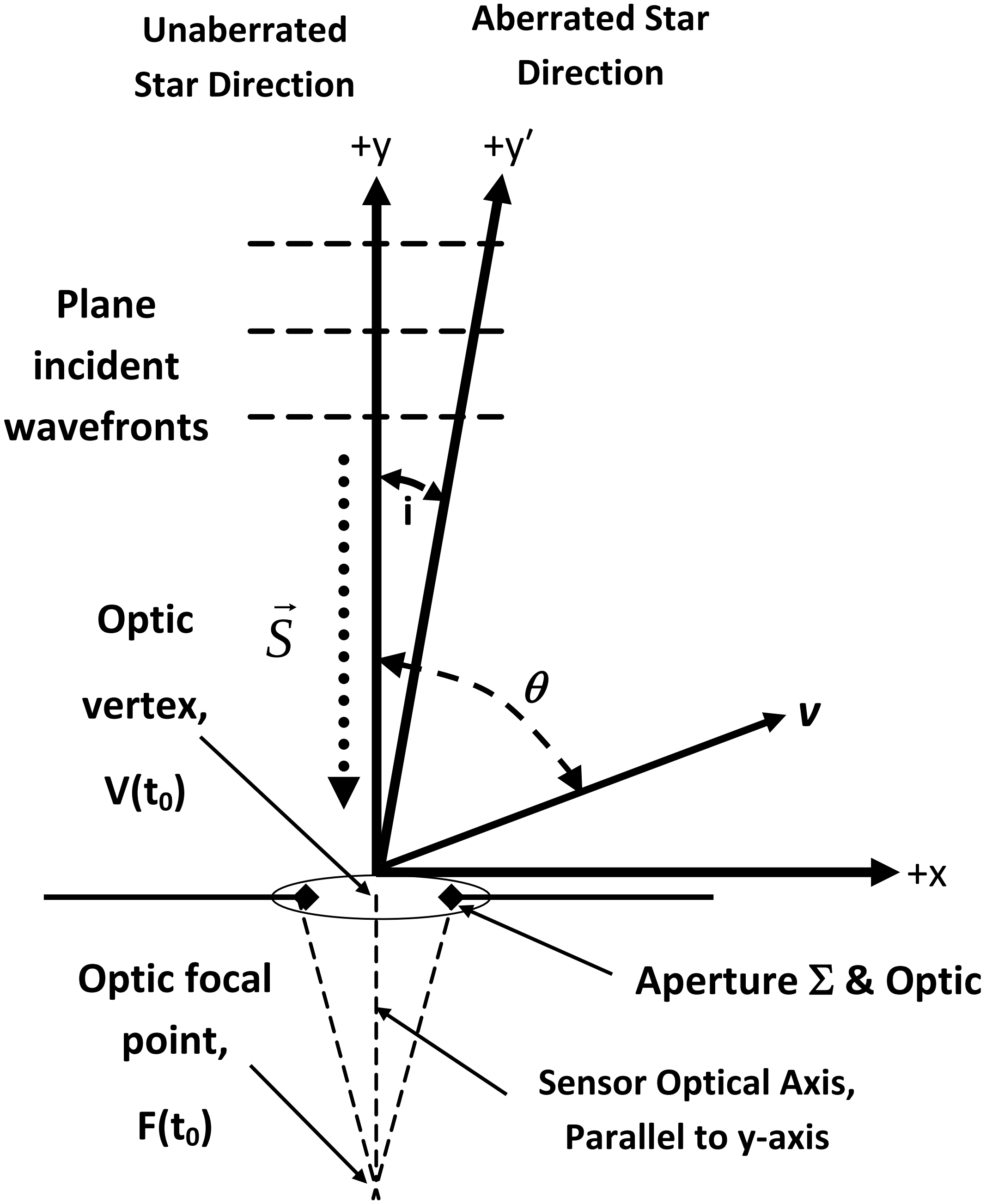

**Figure 1**: Overall geometry. +y axis and sensor Optical Axis parallel to line of sight to Unaberrated Viewing Direction to star. Plane incident wavefronts from star are perpendicular to y-axis. Their Poynting Vector, $\vec{S}$ , is parallel to y-axis, and positive in –y direction. Angle between y-axis and sensor velocity direction, $\vec{v}$ , is $\theta$ . Aberrated Viewing Direction is $+\ y'$ axis in x/y plane. Aberration angle is $i$ . Both $\vec{v}$ and $i$ lie in x/y plane. Shape Σ defines optics aperture. Optic images star in sensor focal plane. Sensor velocity component parallel to incident planar wavefronts is $v_x = v\sin(\theta)$ .

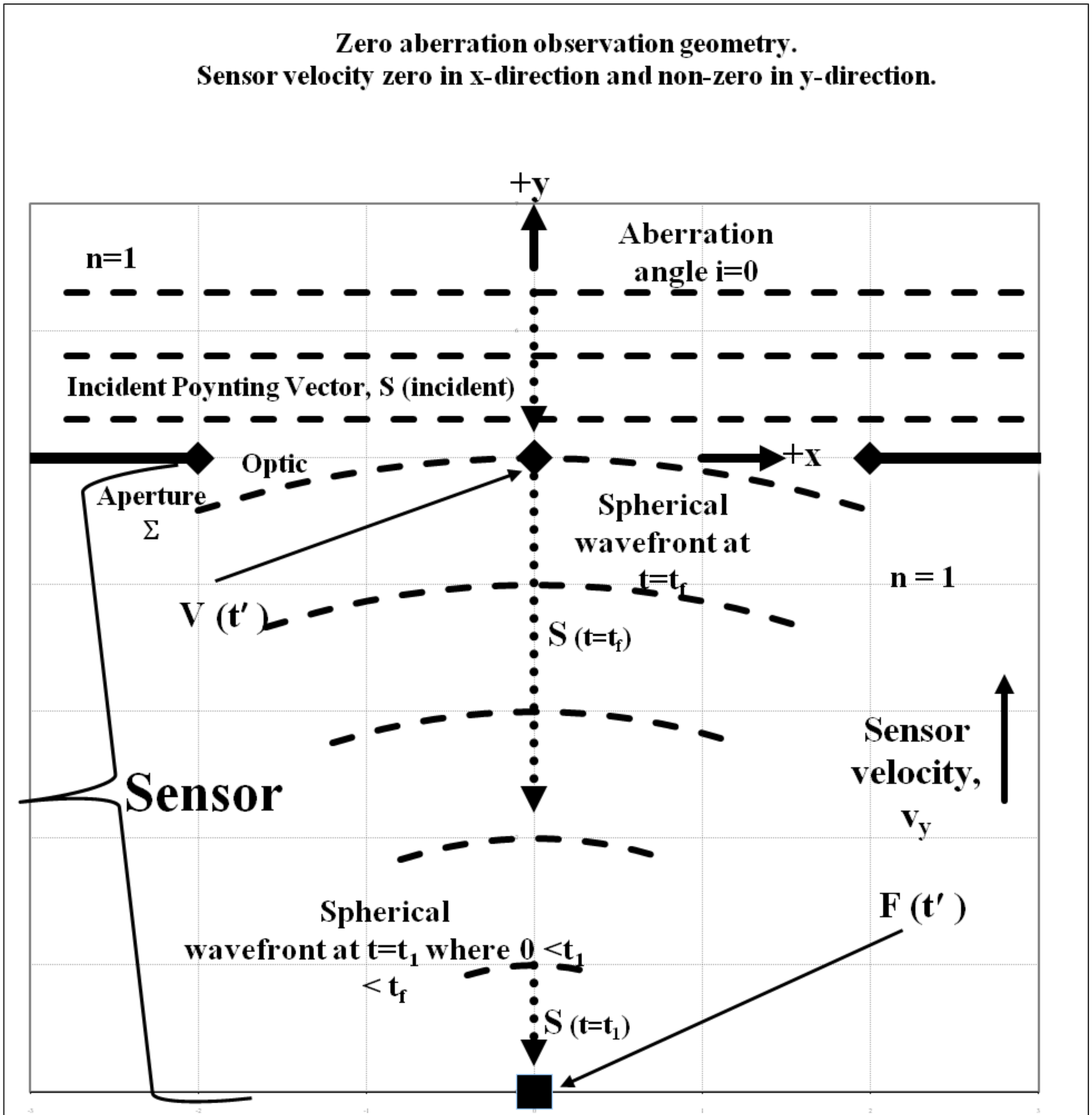

**Figure 2**: *θ=0* observation geometry in reference frame of sensor. Sensor velocity parallel to Unaberrated Viewing Direction. Sensor optical axis is line containing V(t′) and F(t′) for arbitrary time t′. Optical axis is parallel to sensor velocity. Sensor views star at zero aberration independent of source motion.

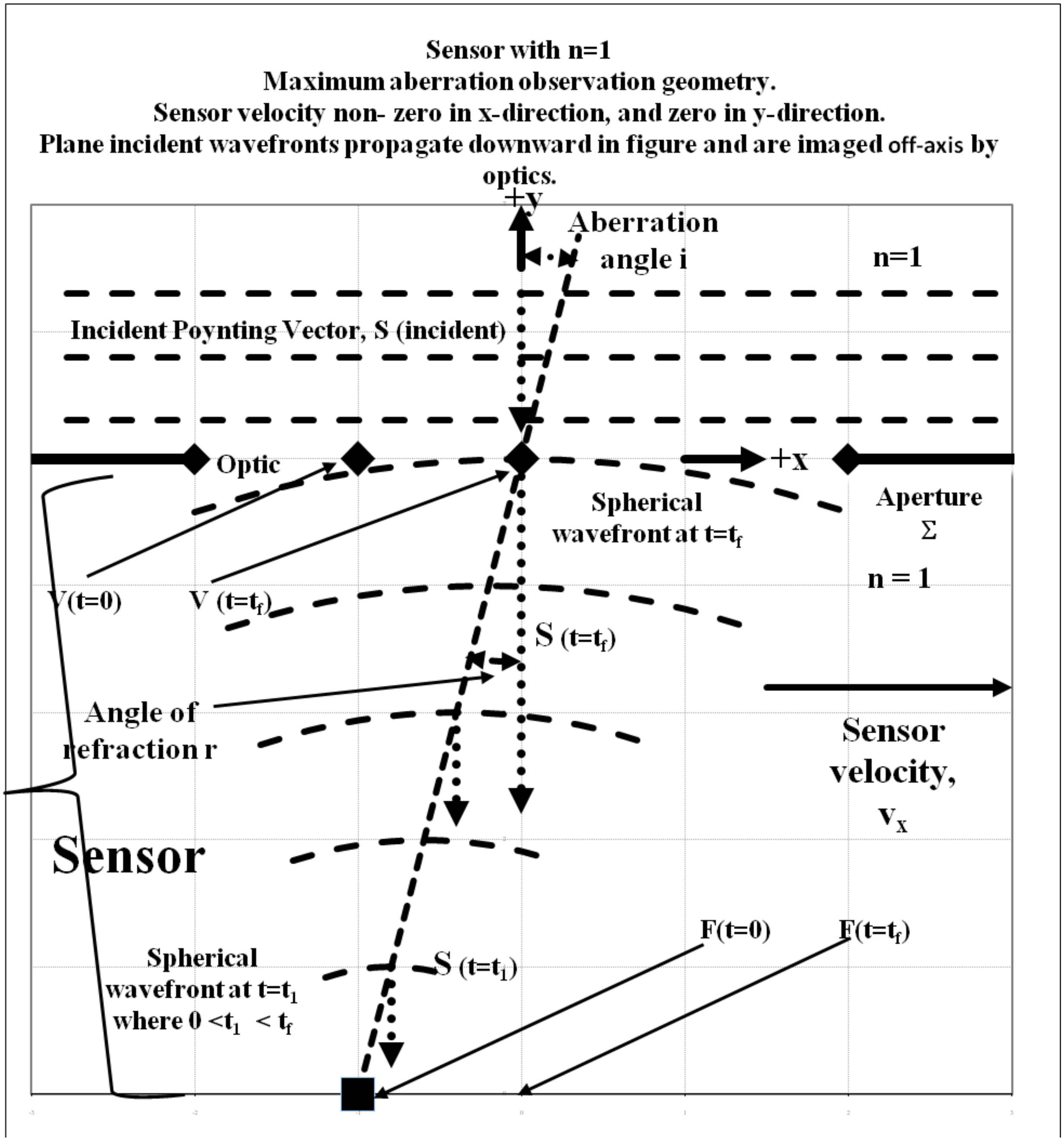

**Figure 3**: In the reference frame of moving sensor. Vacuum filled viewing stellar aberration at maximum magnitude where sensor velocity is normal to Unaberrated Viewing Direction. The wavefronts shear parallel to the direction of sensor motion as they converge to form an image of the star. The wavefront shear before the lens has no effect since the incident wavefronts are plano and shear in plane, so aberration is independent of source motion.

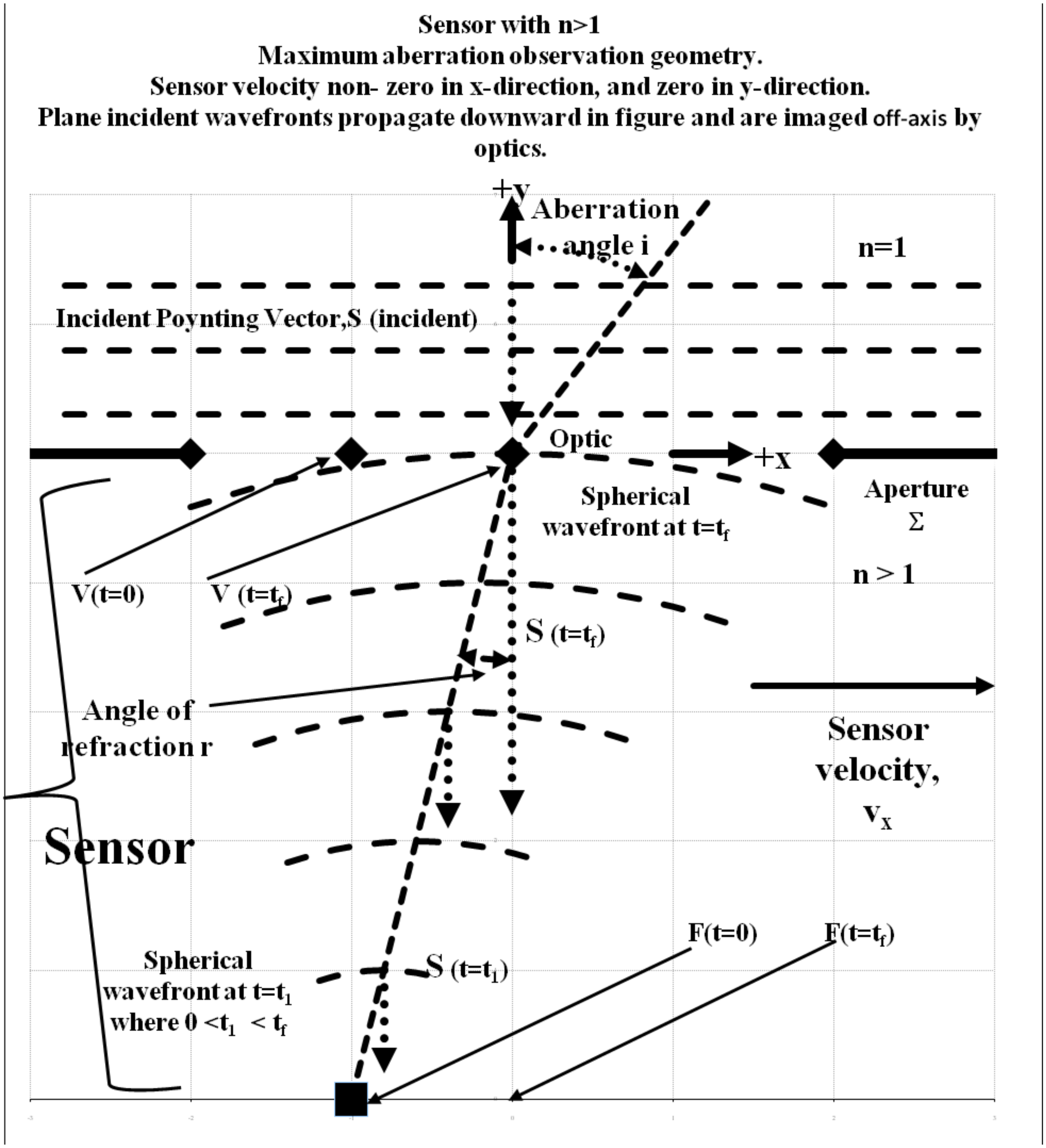

**Figure 4**: In frame of reference of moving sensor. Sensor, filled with non-vacuum, viewing stellar aberration at maximum magnitude. Sensor velocity is normal to Unaberrated Viewing Direction. Aberration depends on index of refraction within sensor. Snell's law on the center ray increases the aberration angle to *i* from *r*, $\sin(i) = n \sin(r)$ where $\tan(r) = nv/c$.

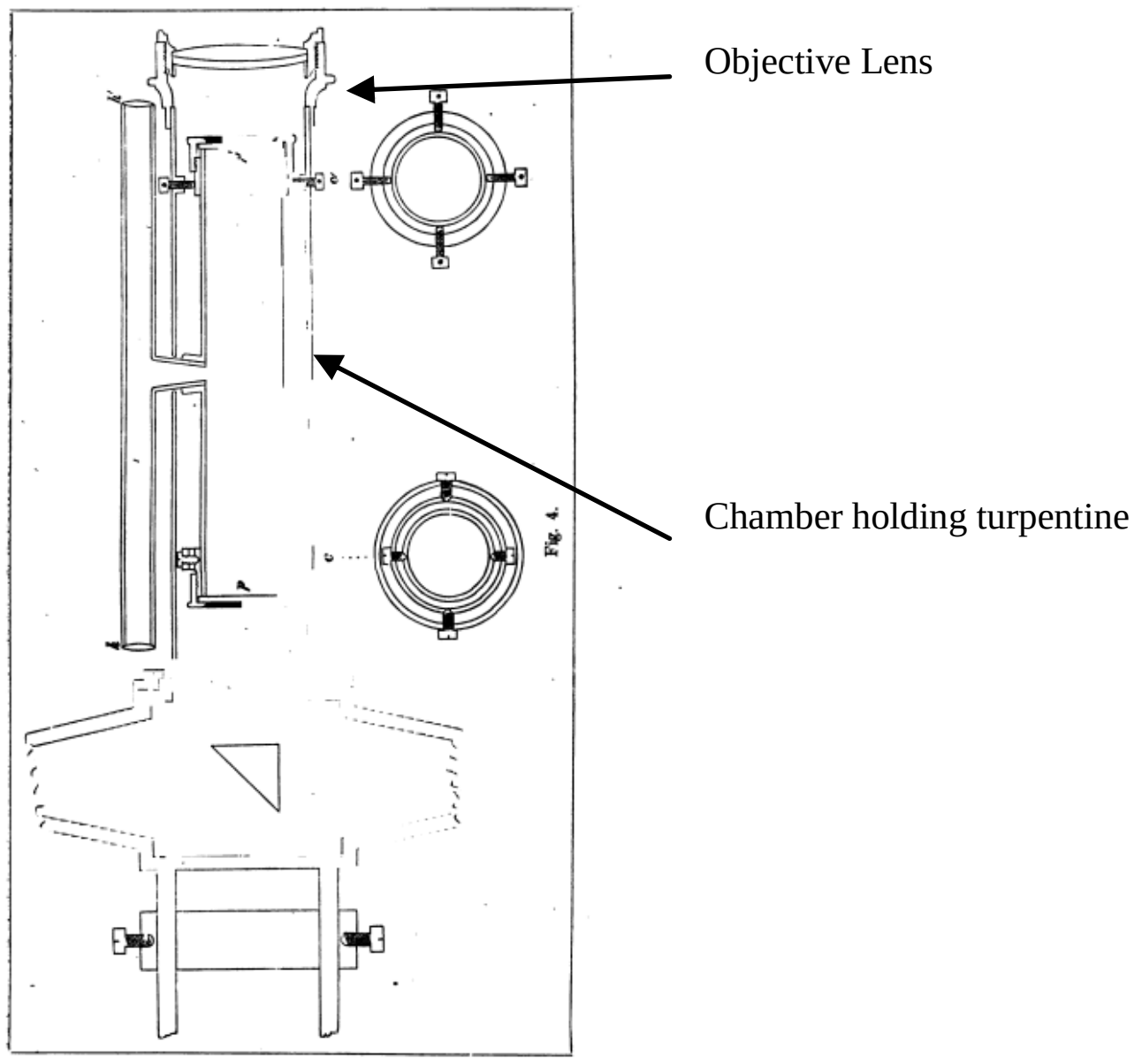

**Figure 5**

Klinkerfues's sensor with glass-enclosed cavity for liquid column [18]